\newcommand{\GCD}{\mbox{GCD}}
\newcommand{\EGCD}{\mbox{EGCD}}

\documentclass[11pt]{article}
\usepackage{amsmath}
\newcommand{\shorttitletext}{Hing Leung}

\usepackage{amsmath}
\begin{document}

\sloppy
{\large\bf
{\center 
A Note on 
Extended Euclid's Algorithm\\
}
}


{\center Hing Leung\\
Department of Computer Science\\
New Mexico State University\\
Las Cruces, NM 88003, USA\\
}

\begin{abstract}
Starting with the recursive extended Euclid's algorithm, 
we apply a systematic approach using matrix notation
to transform it into an iterative algorithm.
The partial correctness proof derived from the transformation
turns out to be very elegant, 
and easy to follow.
The paper provides a connection between recursive and iterative
versions of extended Euclid's algorthm.

\end{abstract}

\section{Introduction}

In teaching computer algorithms, many textbooks 
favor recursive over iterative versions.
While the correctness of recursive algorithms
can be relatively easy to establish, students may sometimes find it difficult to
connect it to the more efficient iterative versions.
An example is the extended Euclid's algorithm from which we compute
multiplicative inverses in modular arithmetic.
In the textbooks,
either the recursive~\cite{Cormen, Dasgupta, Epp, Goodrich}
or iterative~\cite{Aho, Bach, Knuth1, Knuth2, Koshy}
version is presented, but not both.
Moreover, the recursive and iterative versions of extended Euclid's algorithm
look drastically different. 
Computing multiplicative inverses efficiently is of
great 
importance as it is an indispensible component in the implementation of 
RSA cryptosystem. As 
the numbers involved are of hundred of bits in length,
it is without doubt that a recursive
algorithm may be unacceptably slow.


In this paper,
starting with the recursive extended Euclid's algorithm, 
we apply a systematic approach to transform it into an iterative algorithm. 
Using matrix notation, and making use of the associative
property of matrix multiplication, one obtains an iterative algorithm. 
The partial correctness proof turns out to be very clean
with the loop invariant expressed using matrices.
The paper thus provides a connection between recursive and iterative
versions of extended Euclid's algorthm.

Recursive versions of extended Euclid's algorithm
are provided in textbooks~\cite{Cormen, Dasgupta, Epp, Goodrich}.
The textbook~\cite{Epp}
by Epp 
provides a detailed account of the recursive ideas, 
supplemented with demonstrations by hand calculations.
In~\cite{Aho}, an iterative extended Euclid's algorithm is presented
while the recursive version is missing. 
Using mathematical induction, the correctness proof
of the iterative algorithm is established in~\cite{Aho} through insightful observation of the properties
of the numbers computed during the execution of the algorithm. 
Then in~\cite{Bach}, matrices
are used\footnote{
In fact, matrices are introduced in~\cite{Aho}, 
not for the presentation of the extended Euclid's algorithm, 
only for the development of an algorithm for computing the greatest
common divisor for polynomials over a field.
} 
to present the extended Euclid's algorithm iteratively. 
However, the development of the iterative codes requires good insights 
in mathematics, and does not originate from the recursive algorithm,
which the book does not give either. 
Also, the way the iterative codes are given in~\cite{Bach} 
is not suitable for the construction of a partial correctness proof.
Knuth establishes in~\cite{Knuth1} (Section 1.2.1) the partial correctness proof of 
the iterative extended Euclid's algorithm
\cite{Knuth1, Knuth2} through 
a flow chart program labeled with assertions.
In fact, the non-trivial correctness proof was the only 
example that Knuth gives in Chapter 1 
on basic concepts of his classic book~\cite{Knuth1} 
for demonstrating how partial correctness proof can be performed. Even though Knuth's proof
is not difficult to follow,
it is still quite mysterious with no indication of how 
the loop invariant 
is derived or developed. In contrast, our partial correctness proof is 
logically developed and step-by-step justified;
and the proof's succinctness and elegantness are appealing, not intimidating.
In this paper, we do not consider variants \cite{Gordon, Knuth2, Stein}
of iterative extended Euclid's algorithms for improved efficiency.

\section{Recursive Extended Euclid's Algorithm}

We give in this section a quick 
review of the design
of the recursive extended Euclid's algorithm~\cite{Cormen, Dasgupta, Epp, Goodrich}.

Below is Euclid's algorithm for computing the greatest common divisor of $a$ and $b$, 
where $a \geq b \geq 0$: 

\newpage

{\tt\bf\footnotesize
GCD(a,b):
\hspace{0.0in} {// assume a $\geq$ b $\geq$ 0}\\
\hspace*{0.2in} if (b == 0) return a\\
\hspace*{0.2in} return GCD(b,a\%b)
}

Euclid's algorithm can be extended to return integers $x$ and $y$ 
such that $\GCD(a,b) = ax + by$. One can develop the extended Euclid's algorithm
by first taking a leap of faith that such an algorithm $\EGCD(a,b)$ can be constructed 
by extending $\GCD(a,b)$ to
return $(d,x,y)$ where $d = $GCD$(a,b)$ and $d = ax + by$. The algorithm
is thus organized as follows:

{\tt\bf\footnotesize
EGCD(a,b):                                              \hspace{0.0in} { // assume a $\geq$ b $\geq$ 0}\\
\hspace*{0.2in} if (b == 0) return (a,1,0)       \hspace{0.35in} { //  a =  a 1 + b 0}
\hspace{0.608in} {\footnotesize  (base case)}\\
\hspace*{0.2in}  (d,x${\bf '}$,y${\bf '}$) = EGCD(b,a\%b) \hspace{0.41in}  //  d =  b x${\bf '}$ + (a \% b) y${\bf '}$
\hspace{0.1in} {\footnotesize (hypothesis)}\\
\hspace*{0.2in} /* steps are needed to compute x and y */  \\
\hspace*{0.22in} return  (d,x,y) \hspace{1.15in} //  d =  a x +  b y \hspace{0.6in} {\footnotesize (goal)}
}

We need to determine the steps to compute $x$ and $y$.  
From the hypothesis, we deduce that
$d = b x' + (a \% b) y'$, which can be rewritten as $d = b x' + (a - \lfloor a/b \rfloor b) y' = a y' + b (x' - \lfloor a/b \rfloor y')$.
By letting $x = y'$ and $y = x' - \lfloor a/b \rfloor y'$,
the algorithm is finalized as follows:

{\tt\bf\footnotesize
EGCD(a,b):                                              \hspace{0.0in} { // assume a $\geq$ b $\geq$ 0}\\
\hspace*{0.2in} if (b == 0) return (a,1,0)       \hspace{0.35in} { //  a =  a 1 + b 0}
\hspace{0.608in} {\footnotesize  (base case)}\\
\hspace*{0.2in}  (d,x${\bf '}$,y${\bf '}$) = EGCD(b,a\%b) \hspace{0.41in}  //  d =  b x${\bf '}$ + (a \% b) y${\bf '}$
\hspace{0.1in} {\footnotesize (hypothesis)}\\
\hspace*{0.21in}  (x,y)  = (y${\bf '}$,x${\bf '}$-$\lfloor$a/b$\rfloor$y${\bf '}$) \\
\hspace*{0.22in} return  (d,x,y) \hspace{1.15in} //  d =  a x +  b y \hspace{0.61in} {\footnotesize (goal)}
}

Rewriting $a \% b$ as $a - \lfloor a/b \rfloor b$,
the algorithm can be given as follows:

{\tt\bf\footnotesize
EGCD(a,b):                                              \hspace{0.0in} { // assume a $\geq$ b $\geq$ 0}\\
\hspace*{0.2in} if (b == 0) return (a,1,0)       \hspace{0.35in} { //~a =  a 1 + b 0}
\hspace{0.60in} {\footnotesize  (base case)}\\
\hspace*{0.2in}  (d,x${\bf '}$,y${\bf '}$) = EGCD(b,a - $\lfloor$a/b$\rfloor$b) 
\hspace{0.15in}  //  d =  b x${\bf '}$ + (a \% b) y${\bf '}$
\hspace{0.25in} {\footnotesize (hypothesis)}\\
\hspace*{0.21in}  (x,y)  = (y${\bf '}$,x${\bf '}$-$\lfloor$a/b$\rfloor$y${\bf '}$) \\
\hspace*{0.22in} return  (d,x,y) \hspace{1.18in} //  d =  a x +  b y \hspace{0.57in} {\footnotesize (goal)}
}

\section{From Recursion to Iteration}


We notice the similariy between 
the computing of $(y',x'-\lfloor a/b \rfloor y')$ from $(x',y')$
and
the computing of $(b,a-\lfloor a/b \rfloor b)$ from $(a,b)$.
We can capture the computation using matrix multiplication.
Let
{\footnotesize $A = 
\begin{bmatrix}
0 & 1 \\
1 & -\lfloor a/b \rfloor 
\end{bmatrix}
$}.
Then
$
[x~y]
= 
[x'~y']
A 
$
and
$
[b ~~
a-\lfloor a/b \rfloor b]
= [a~b]
A 
$.

Our goal is to derive an iterative extended Euclid's algorithm EGCD. 
But, for technical convenience, for the moment we reduce the functionality of EGCD to
eGCD, which performs the same as before except that 
it no longer returns $d$, the greatest common divisor of $a$ and $b$.
Furthermore, eGCD takes vectors for both input and output.
Using matrix notation, we derive the codes for eGCD($[a~b]$) 
which returns $[x~y]$ such that $\gcd(a,b) = ax + by$ 
as follows:
\vspace*{0.1in}

{\tt\bf\footnotesize
eGCD([a~b]):                                              \hspace{0.0in} { // assume a $\geq$ b $\geq$ 0}\\ \\
\hspace*{0.2in} if (b == 0) return [1 0]
\\  \\
\hspace*{0.2in} return  
eGCD 
$\left( [\textbf{a} ~ \textbf{b}] 
\begin{bmatrix}
\textbf{0} & \textbf{1} \\
\textbf{1} & -\lfloor \textbf{a}/\textbf{b} \rfloor
\end{bmatrix}
\right)
$
$
\begin{bmatrix}
\textbf{0} & \textbf{1} \\
\textbf{1} & -\lfloor \textbf{a}/\textbf{b} \rfloor 
\end{bmatrix}
$
}

As the recursion progresses, the recursion stack remembers the sequence of 2$\times$2 matrices generated. 
It is only when the recursion halts 
that we start multiplying the result $[1~0]$
with the sequence of matrices in the reverse order of their generations.
However, since matrix multiplication is associative, we can indeed multiply the
2$\times$2 matrices as they are generated. That is, we no longer need to remember
all the matrices generated until the program halts.
This insight allows us to derive an iterative algorithm.

Each time the program advances to the next round, the matrix 
{\footnotesize
$
\begin{bmatrix}
0 & 1 \\
1 & -\lfloor a/b \rfloor
\end{bmatrix}
$
}
is multiplied to the left side of the product of 2$\times$2 matrices
generated so far. 
In addition,
{\footnotesize
$
\begin{bmatrix}
0 & 1 \\
1 & -\lfloor a/b \rfloor
\end{bmatrix}
$
}
is multiplied to the right side of $[a~b]$ to give the new $[a~b]$.
By maintaining the product of 2$\times$2 matrices generated as 
$
\begin{bmatrix}
c & e \\
d & f 
\end{bmatrix}
$,
the codes for eGCD($[a,b]$) becomes:

\newpage

{\tt\bf\footnotesize
eGCD([a  b]): 
\\
\\
\hspace*{0.2in} 
$
\begin{bmatrix}
\textbf{c} & \textbf{e} \\
\textbf{d} & \textbf{f} 
\end{bmatrix}
$
=
$
\begin{bmatrix}
\textbf{1} & \textbf{0} \\
\textbf{0} & \textbf{1} 
\end{bmatrix}
$
\\
\\
\hspace*{0.2in}
{\bf
   while ($\textbf{b}$ != \textbf{0})
} 
\\
\\
\hspace*{0.4in}
$
\begin{bmatrix}
\textbf{c} & \textbf{e} \\
\textbf{d} & \textbf{f} 
\end{bmatrix}
$
=
$
\begin{bmatrix}
\textbf{0} & \textbf{1} \\
\textbf{1} & - \lfloor \textbf{a}/\textbf{b} \rfloor
\end{bmatrix}
$
$
\begin{bmatrix}
\textbf{c} & \textbf{e} \\
\textbf{d} & \textbf{f} 
\end{bmatrix}
$
\\ \\
\hspace*{0.4in}
$
[\textbf{a} ~ \textbf{b}]
$
=
$
[\textbf{a} ~ \textbf{b}]
$
$
\begin{bmatrix}
\textbf{0} & \textbf{1} \\
\textbf{1} & - \lfloor \textbf{a}/\textbf{b} \rfloor
\end{bmatrix}
$
\\ \\
\hspace*{0.2in}
{\bf return}
$[\textbf{1}~\textbf{0}] 
\begin{bmatrix}
\textbf{c} & \textbf{e} \\
\textbf{d} & \textbf{f} 
\end{bmatrix}
$
}

As the matrix
$
\begin{bmatrix}
0 & 1 \\
1 & - \lfloor a/b \rfloor
\end{bmatrix}
$
is the transpose of itself, we have
$$
\begin{bmatrix}
c & d \\
e & f
\end{bmatrix}
=
\begin{bmatrix}
c & d \\
e & f
\end{bmatrix}
\begin{bmatrix}
0 & 1 \\
1 & - \lfloor a/b \rfloor
\end{bmatrix}
$$
Next, we give the codes of eGCD($[a,b]$) in a compact way as follows:

{\tt\bf\footnotesize
eGCD([a, b]): 
\\ \\
\hspace*{0.2in} 
$
\begin{bmatrix}
\textbf{c} & \textbf{d} \\
\textbf{e} & \textbf{f} 
\end{bmatrix}
$
=
$
\begin{bmatrix}
\textbf{1} & \textbf{0} \\
\textbf{0} & \textbf{1} 
\end{bmatrix}
$
\\ \\
\hspace*{0.2in}
{\bf
   while ($\textbf{b}$ != \textbf{0})
} 
\\ \\
\hspace*{0.4in}
$
\begin{bmatrix}
\textbf{a} & \textbf{b} \\
\textbf{c} & \textbf{d} \\
\textbf{e} & \textbf{f} 
\end{bmatrix}
$
=
$
\begin{bmatrix}
\textbf{a} & \textbf{b} \\
\textbf{c} & \textbf{d} \\
\textbf{e} & \textbf{f} 
\end{bmatrix}
$
$
\begin{bmatrix}
\textbf{0} & \textbf{1} \\
\textbf{1} & - \lfloor \textbf{a}/\textbf{b} \rfloor
\end{bmatrix}
$
\\ \\
\hspace*{0.2in}
{\bf return}
$
[\textbf{c} , \textbf{e}]$
}

By the logic of the original Euclid's algorithm, the greatest common divisor is given by
the value of $a$ when the program GCD halts. 
It turns out that the same is true for the program eGCD.
We therefore obtain iterative codes for EGCD by augmenting eGCD to return $(a,c,e)$.
Furthermore, the current codes of eGCD are treating the input 
parameters $a$ and $b$ as working variables, which values change over the execution of
the program.
By separating the input parameters from the working variables,
we obtain the iterative EGCD algorithm as follows:

\newpage
{\tt\bf\footnotesize
EGCD($\boldsymbol{\alpha}, \boldsymbol{\beta}$): 
\\ \\
\hspace*{0.2in} 
$
\begin{bmatrix}
\textbf{a} & \textbf{b} \\
\textbf{c} & \textbf{d} \\
\textbf{e} & \textbf{f} 
\end{bmatrix}
$
=
$
\begin{bmatrix}
\boldsymbol{\alpha} & \boldsymbol{\beta} \\
\textbf{1} & \textbf{0} \\
\textbf{0} & \textbf{1} 
\end{bmatrix}
$
\\ \\
\hspace*{0.2in}
{\bf
   while ($\textbf{b}$ != \textbf{0})
} 
\\ \\
\hspace*{0.4in}
$
\begin{bmatrix}
\textbf{a} & \textbf{b} \\
\textbf{c} & \textbf{d} \\
\textbf{e} & \textbf{f} 
\end{bmatrix}
$
=
$
\begin{bmatrix}
\textbf{a} & \textbf{b} \\
\textbf{c} & \textbf{d} \\
\textbf{e} & \textbf{f} 
\end{bmatrix}
$
$
\begin{bmatrix}
\textbf{0} & \textbf{1} \\
\textbf{1} & - \lfloor \textbf{a}/\textbf{b} \rfloor
\end{bmatrix}
$
\\ \\
\hspace*{0.2in}
{\bf return}
$
(\textbf{a}, \textbf{c} , \textbf{e} )$
}

Note that the codes above are the same as that given in Knuth's Algorithm X in page 342 of~\cite{Knuth2} except that we are using matrix notations whereas Kunth uses vectors.

\section{Partial Correctness Proof}



In this section, we prove the partial correctness of the iterative EGCD algorithm. First, we 
need to determine the loop invariant.
During the execution of the iterative EGCD algorithm, 
{\footnotesize
$
\begin{bmatrix}
c & d \\
e & f 
\end{bmatrix}
$} 
is the product of the 2$\times$2 matrices
of the form 
$
\begin{bmatrix}
0 & 1 \\
1 & - \lfloor a/b \rfloor
\end{bmatrix}
$
generated so far in the while loop. On the other hand, $[a~b]$ is
$[\alpha~\beta]$ multiplied with the same sequence of 2$\times$2 matrices.
Thus, $[a~b] = [\alpha ~ \beta]$
{\footnotesize
$
\begin{bmatrix}
c & d \\
e & f 
\end{bmatrix}
$}.
Next, since $[a~b]$ 
{\footnotesize
$
\begin{bmatrix}
0 & 1 \\
1 & - \lfloor a/b \rfloor
\end{bmatrix}
$
}
= $[b~a\%b]$ and recall that \GCD($a,b$) = \GCD($b,a\%b$)
by the logic of Euclid's algorithm,
the greatest common divisor 
of the current values of $a$ and $b$ must equal 
the greatest common divisor 
of the previous values of $a$ and $b$, which in turn must equal
the greatest common divisor 
of the initial values of $a$ and $b$, which are $\alpha$ and $\beta$ respectively.
Therefore, $\GCD(a,b) = \GCD(\alpha,\beta)$.

We propose the loop invariant to consist of the 
two relationship
$$[a~b] = [\alpha~\beta]
\begin{bmatrix}
c & d \\
e & f 
\end{bmatrix}~\textrm{and}~ \GCD(a,b) = \GCD(\alpha,\beta).$$

Next, we will complete the partial correctness proof step by step.

We need to verify that the the loop invariant holds when the while loop is first entered.
After the initial assignment statement 
before the while loop is executed,
we have
$[a~b] = [\alpha~\beta]$ and
{\footnotesize
$
\begin{bmatrix}
c & d \\
e & f 
\end{bmatrix}
$
}
=
{\footnotesize
$
\begin{bmatrix}
1 & 0 \\
0 & 1 
\end{bmatrix}
$}, for which the loop invariant can be easily verified to be valid.

We now turn the attention to the assignment statement within the while loop.
The precondition of the assignment statement is the loop invariant 
augmented with $b \neq 0$. 
Even though the assignment statement involves division by $b$, 
it is a safe operation as the precondition guarantees that  $b$ is non-zero.

From 
$[a~b] = [\alpha~\beta]
\begin{bmatrix}
c & d \\
e & f 
\end{bmatrix}$ of the loop invariant, 
we can preserve equality by multiplying to each side of the equation the matrix 
{\footnotesize
$
\begin{bmatrix}
0 & 1 \\
1 & - \lfloor a/b \rfloor
\end{bmatrix}
$
}
as $b \neq 0$.
Together with the facts that $\GCD(b,a\%b) = \GCD(a,b)$ when $b \neq 0$ (by the 
logic of Euclid's algorithm) and $\GCD(a,b) = \GCD(\alpha,\beta)$ (by the loop invariant),
we deduce that the following holds:
$$
[a~b]  
\begin{bmatrix}
0 & 1 \\
1 & - \lfloor a/b \rfloor
\end{bmatrix}
= [\alpha~\beta]~
\begin{bmatrix}
c & d \\
e & f 
\end{bmatrix}
\begin{bmatrix}
0 & 1 \\
1 & - \lfloor a/b \rfloor
\end{bmatrix},~\textrm{and}$$
$$ \GCD(b,a\%b) = \GCD(a,b) = \GCD(\alpha,\beta),~~\textrm{and}~~b \neq 0 $$

After performing the assignment statement, 
the value $[a~b]  
\begin{bmatrix}
0 & 1 \\
1 & - \lfloor a/b \rfloor
\end{bmatrix}
$ is re-written as the new $[a~b]$; 
$
\begin{bmatrix}
c & d \\
e & f 
\end{bmatrix}
\begin{bmatrix}
0 & 1 \\
1 & - \lfloor a/b \rfloor
\end{bmatrix}
$
is re-written as the new
$
\begin{bmatrix}
c & d \\
e & f 
\end{bmatrix}
$; $\GCD(b,a\%b)$ is rewritten as $\GCD(a,b)$ since
the new $[a~b]$ has taken on the value  $[a~b] 
\begin{bmatrix}
0 & 1 \\
1 & - \lfloor a/b \rfloor
\end{bmatrix}
= [b~a\%b]$;
however, the value of $a \% b$, which becomes the new $b$ value,
can no longer be guaranteed to be non-zero.
Putting the observations together, the loop invariant again holds after the assignment statement.

Finally, we consider the correctness when the loop is exited.
Upon exiting the while loop when $b = 0$, the loop invariant still holds. 
Instantiating $b$ to $0$ in the condition $\GCD(a,b) = \GCD(\alpha,\beta)$
of the loop invariant, we have $a = \GCD(a,0) = \GCD(a,b) = \GCD(\alpha,\beta)$. 
From the condition 
$[a~b] = [\alpha~\beta]~
\begin{bmatrix}
c & d \\
e & f 
\end{bmatrix}$
of the loop invariant,
we deduce that $a = \alpha c + \beta e$. 
Therefore, the return of the triple $(a,c,e)$ by EGCD is correct
as $a =  \GCD(\alpha,\beta)$.

\section{Conclusion}

In this paper,
using the leap of faith technique, the recursive extended Euclid's algorithm is first constructed.
Taking the original Euclid's algorithm for granted, we do not need any new mathematics insights 
in extending Euclid's algorithm.
Next, applying only simple program 
transformation technique, and using the associative property of matrix multiplication, 
we derive the iterative algorithm. The clean process easily pinpoints the loop invariant from which  
a solid and easy-to-follow partial correctness proof of the iterative algorithm is 
established.

The treatment given in the paper demonstrates the maturity of computer science as a discipline.
Relying only on computer science principles, we successfully derive the iterative
extended Euclid's algorithm. Our approach complements other developments 
of the iterative algorithm~\cite{Aho, Bach, Knuth1, Knuth2, Koshy} that require more mathematics insights
in their presentations.

\end{document}